\documentclass[twoside,english,british,aip,apl,reprint]{revtex4-1}
\usepackage[T1]{fontenc}
\usepackage[latin9]{inputenc}
\usepackage{float}
\usepackage{units}
\usepackage{amsmath}
\usepackage{amssymb}
\usepackage{graphicx}



\usepackage{babel}
\begin{document}

\title{Origin and Evolution of Surface Spin Current in Topological Insulators}

\author{André Dankert}
\email{andre.dankert@chalmers.se}

\selectlanguage{british}%

\affiliation{Department of Microtechnology and Nanoscience, Chalmers University
of Technology, Quantum Device Laboratory; Göteborg, Sweden}

\author{Priyamvada Bhaskar}

\affiliation{Department of Microtechnology and Nanoscience, Chalmers University
of Technology, Quantum Device Laboratory; Göteborg, Sweden}

\author{Dmitrii Khokhriakov}

\affiliation{Department of Microtechnology and Nanoscience, Chalmers University
of Technology, Quantum Device Laboratory; Göteborg, Sweden}

\author{Isabel H. Rodrigues}

\affiliation{Department of Microtechnology and Nanoscience, Chalmers University
of Technology, Quantum Device Laboratory; Göteborg, Sweden}

\author{Bogdan Karpiak}

\affiliation{Department of Microtechnology and Nanoscience, Chalmers University
of Technology, Quantum Device Laboratory; Göteborg, Sweden}

\author{M.Venkata Kamalakar}

\affiliation{Department of Microtechnology and Nanoscience, Chalmers University
of Technology, Quantum Device Laboratory; Göteborg, Sweden}

\affiliation{Department of Physics and Astronomy, Uppsala University, Box 516,
SE-75120, Uppsala, Sweden}

\author{Sophie Charpentier}

\affiliation{Department of Microtechnology and Nanoscience, Chalmers University
of Technology, Quantum Device Laboratory; Göteborg, Sweden}

\author{Ion Garate}

\affiliation{Département de Physique, Institut Quantique and Regroupement Québécois
sur les Matériaux de Pointe, Université de Sherbrooke, Sherbrooke,
Québec, Canada J1K 2R1 }

\author{Saroj P. Dash}
\email{saroj.dash@chalmers.se}

\selectlanguage{british}%

\affiliation{Department of Microtechnology and Nanoscience, Chalmers University
of Technology, Quantum Device Laboratory; Göteborg, Sweden}
\begin{abstract}
The Dirac surface states of topological insulators offer a unique
possibility for creating spin polarized charge currents due to the
spin-momentum locking. Here we demonstrate that the control over the
bulk and surface contribution is crucial to maximize the charge-to-spin
conversion efficiency. We observe an enhancement of the spin signal
due to surface-dominated spin polarization while freezing-out the
bulk conductivity in semiconducting Bi$_{1.5}$Sb$_{0.5}$Te$_{1.7}$Se$_{1.3}$
below $\unit[100]{K}$. Detailed measurements up to room temperature
exhibit a strong reduction of the magnetoresistance signal between
$2-\unit[100]{K}$, which we attribute to the thermal excitation of
bulk carriers and to the electron-phonon coupling in the surface states.
The presence and dominance of this effect up to room temperature is
promising for spintronic science and technology. 
\end{abstract}


\maketitle
\begin{figure}
\begin{centering}
\includegraphics[scale=1.1]{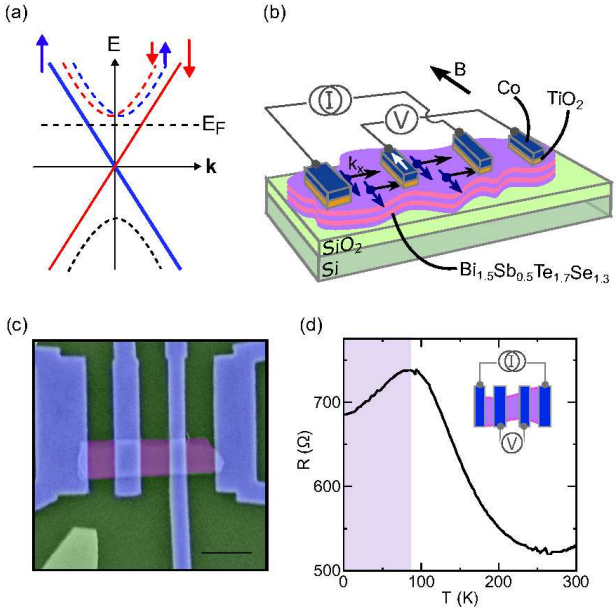}
\par\end{centering}
\caption{\textbf{Device schematic and electrical characterization.} (a)~Dirac
cone of the topological surface state with spin-momentum locking~(SML)
and spin polarized non-topological surface states (dashed lines).
(b)~Schematic of a TI with FM tunnel contacts. The direction of spin
current in Bi$_{1.5}$Sb$_{0.5}$Te$_{1.7}$Se$_{1.3}$ (BSTS) is
defined by the charge current direction due to SML. (c)~Coloured
scanning electron micrograph of a fabricated device with Co/TiO$_{2}$
contacts on an exfoliated BSTS flake ($\unit[1]{\mu m}$ scale bar).
(d)~Temperature dependence of the BSTS channel resistance transitioning
from metallic (coloured background) to semiconducting (white background)
behaviour. Inset: Measurement configuration.  \label{fig:Device-schematic}}
\end{figure}

Three-dimensional (3D) topological insulators (TIs) emerge as a result
of band inversion due to strong spin-orbit (SO) coupling. These band
inversions lead to the appearance of topologically protected gapless
surface states, which have one spin state per momentum at the Fermi
surface (Fig.~\ref{fig:Device-schematic}a).\cite{Xia2009-rv,Hasan2010-am,Pesin2012-uf,Ando2013-zk}
The spins of carriers in TI surface states are locked perpendicular
to their momenta (spin-momentum locking, SML) enabling the creation
of spin polarization by applying a charge current (Fig.~\ref{fig:Device-schematic}b).\cite{Ando2013-zk,Hasan2010-am}
These unique spin polarized surface states of TIs due to SML have
been coupled to ferromagnetic contacts for creating giant spin transfer
effects and for potentiometric detection of the current-induced spin
polarization \cite{Li2014-jw,Ando2014-fh,Tang2014-io,Liu2015-tm,Tian2015-ap,De_Vries2015-np,Li2016-mn,Yang2016-gj}
even up to room temperature.\cite{Dankert2015-ns}
 However, the interpretation of the results has remained challenging
due to mixed contributions from the topological and trivial surface
states on top of a bulk background conduction in prototype TIs such
as Bi$_{2}$Se$_{3}$, Bi$_{2}$Te$_{3}$ and Sb$_{2}$Te$_{3}$,
which have naturally-occurring Se and Te vacancies resulting in unintentional
doping.\cite{Pesin2012-uf} These bulk carriers reduce the current-induced
spin polarization in two ways: (i)~by decreasing the fraction of
the current flowing on the spin-momentum-locked surface states, and
(ii)~by allowing the coupling between the topological surface states
and the non-topological bulk and surface states. In particular, phonon-mediated
surface-bulk coupling can lead to interband scattering between various
spin-orbit states (dashed lines in Fig.~\ref{fig:Device-schematic}a),\cite{Hong2012-zr,Hammar2000-fb,Yazyev2010-kl,Datzer2017-mn}
which affects the detected surface spin polarization. These bulk-related
problems have been improved upon by developing semiconducting bulk
TIs (Bi$_{x}$Sb$_{1-x}$)$_{2}$Te$_{3}$ and Bi$_{1.5}$Sb$_{0.5}$Te$_{1.7}$Se$_{1.3}$~(BSTS),
where the surface state transport dominates at low temperature with
parallel bulk conduction at high temperatures.\cite{Ren2011-gh} However,
only few studies reported the potentiometric detection of SML in such
bulk semiconducting TIs, and were limited to cryogenic temperatures
below $\unit[100]{K}$.\cite{Ando2014-fh,Tang2014-io} The electrical
investigation of the SML phenomenon up to room temperature in such
low doped TIs is crucial for the basic understanding of the influence
of surface state and bulk band contribution to the detected magnetoresistance~(MR)
signal. 

\begin{figure}[t]
\begin{centering}
\includegraphics[scale=1.2]{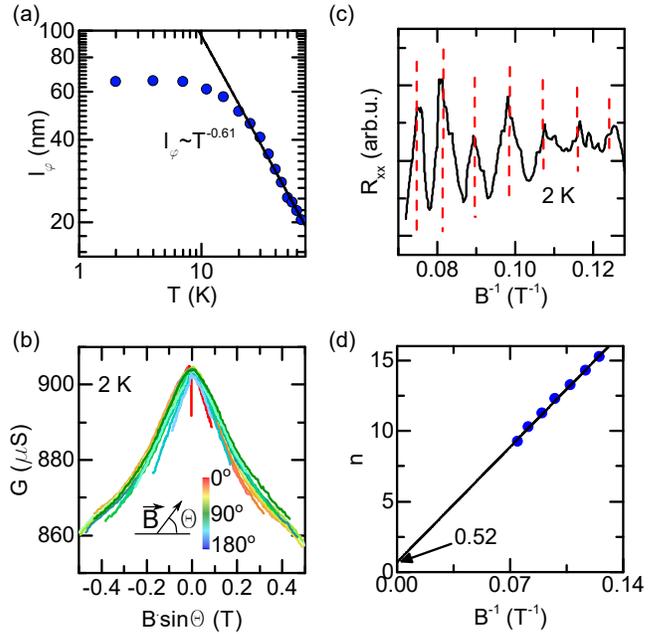}
\par\end{centering}
\caption{\textbf{Weak anti-localization (WAL) and Shubnikov-de~Haas (SdH).}
(a)~Temperature dependence of the phase coherence length. (b)~Angle
dependence of the WAL signal plotted against the out-of-plane magnetic
field component at $\unit[2]{K}$. (c)~Resistance dependence on $B^{-1}$
at $\unit[2]{K}$ measured at $\unit[I=1]{\mu A}$ shows SdH oscillations.
(d)~The position of the oscillation's maxima extrapolates to $n\left(0\right)=\beta=0.52$.\label{fig:WAL-SdH}}
\end{figure}

Here we study the surface spin polarization in BSTS with ferromagnetic
tunnel contacts between $\unit[2-300]{K}$. The semiconducting bulk
nature of BSTS thin films enables us to investigate SML in a surface-dominated
transport regime at low temperatures and a surface-bulk coexistent~(SBC)
conduction regime at high temperatures.  These results provide an
understanding of the origin and evolution of SML signals with temperature
as well as the contribution of bulk bands and topological surface
states.

\textbf{Device and electrical characterization of BSTS.} The devices
were prepared by depositing ferromagnetic~(FM) tunnel contacts (Co/TiO$_{2}$)
on exfoliated BSTS flakes on a SiO$_{2}$/Si substrate (Fig.~\ref{fig:Device-schematic}b
and \ref{fig:Device-schematic}c, as well as Appendix Part~A1).\cite{Dankert2014-ri,Dankert2015-ns}
The tunnelling characteristics of the contacts are presented in Appendix
Fig.~A2. The temperature dependence of BSTS channel resistance in
a flake with a thickness $\unit[t=70]{nm}$ shows an increasing channel
resistance from room temperature to $\unit[90]{K}$, and then decreasing
below $\unit[90]{K}$ (Fig.~\ref{fig:Device-schematic}d). The increase
in resistance when cooling down is expected for semiconducting BSTS
flakes stemming from a parallel bulk contribution, whereas the reduction
in resistance below $\unit[90]{K}$ indicates a freezing out of the
charge carriers in the semiconducting bulk and increasingly dominating
surface transport.\cite{Xia2013-so} The freezing out of the bulk
carrier states in our BSTS is further supported by Hall measurements
showing a steady decrease of the charge carrier concentration with
temperature (see Appendix Fig.~A3). This indicates a remaining surface
transport channel with higher mobility compared to the semiconducting
bulk at low temperature. Using the measured charge carrier concentration
and Fermi energy distribution, we can calculate the Fermi level to
be within the band gap and about $\unit[22]{meV}$ below the conduction
band edge \cite{Kondou2016-nh}.

\begin{figure}
\begin{centering}
\includegraphics[scale=1.2]{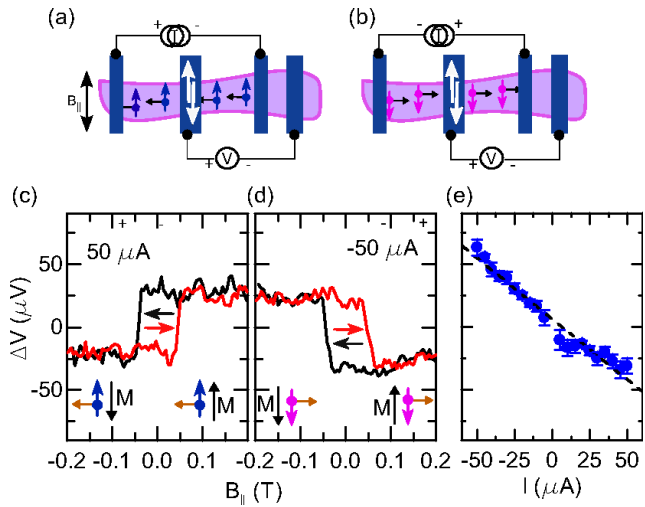}
\par\end{centering}
\caption{\textbf{Electrical detection of SML.} Schematics of the magnetoresistance
measurements between current-induced BSTS surface spins, $S_{\uparrow}$
in (a) and $S_{\downarrow}$ in (b), and magnetization ($\vec{M}$)
of the detector FM contact\cite{Dankert2015-ns}. (c)~Spin signal
shows a hysteretic switching at $\unit[2]{K}$ for a bias current
of $\unit[+50]{\mu A}$. The arrows show the magnetic field sweep
directions up (red) and down (black). (d)~Reversing the current direction
($\unit[-50]{\mu A}$ locking to $S_{\downarrow}$) results in a spin
signal with an inverted hysteretic switching. (e)~Bias current dependence
of the spin signal amplitude $\Delta V=(\mu_{\uparrow}-\mu_{\downarrow})/-e$
measured at $\unit[10]{K}$ for BSTS.\label{fig:SML}}
\end{figure}

\textbf{Magnetotransport and quantum oscillations.} Magnetotransport
measurements in BSTS show a weak anti-localization~(WAL) behaviour
up to $\unit[55]{K}$ indicating a strong SO coupling (Appendix Fig.~A4).
Using the Hikami-Larkin-Nagaoka~(HLN) model, we can fit the conductivity
correction ($\Delta\sigma$) of the surface states with

\begin{equation}
\Delta\sigma\left(B_{\perp}\right)=\alpha\frac{e^{2}}{\pi h}\left[\text{ln}\frac{\hbar}{4el_{\varphi}^{2}B_{\perp}}-\psi\left(\frac{1}{2}+\frac{\hbar}{4el_{\varphi}^{2}B_{\perp}}\right)\right],\label{eq:WAL}
\end{equation}
where $\psi$ represents the digamma function, $l_{\varphi}$ is the
phase coherence length, and $\alpha$ represents the dimensionality
factor of the quantum system\cite{Hikami1980-hi}. The temperature
dependence of the phase coherence length $l_{\varphi}\propto T^{-0.61}$
above $\unit[10]{K}$ (Fig.~\ref{fig:WAL-SdH}a) indicates a 2D system
with dominant phonon scattering ($l_{\varphi}\propto T^{-\nicefrac{1}{2}}$),
which is comparable to previous reports on surface-dominant BSTS \cite{Xia2013-so}
validating the usage of the HLN model.  Figure~\ref{fig:WAL-SdH}b
shows the measured conductivity at $\unit[2]{K}$ for an angle $\varTheta$
from \foreignlanguage{english}{$0^{\circ}$} to \foreignlanguage{english}{$180^{\circ}$}
plotted against the perpendicular magnetic field component $B_{\perp}=B\text{sin}\left(\varTheta\right)$.
The angle independence of this normalized WAL signal together with
its short phase coherence length and its characteristic temperature
dependence confirms the 2D surface states as the dominating transport
channel at cryogenic temperatures.

In high magnetic field measurements, we also observed Shubnikov-de
Haas~(SdH) quantum oscillations in the longitudinal resistance with
a perpendicular magnetic field up to $\unit[14]{T}$ at $\unit[2]{K}$.
Figure~\ref{fig:WAL-SdH}c shows the oscillations as a function of
the inverse magnetic field, where the parabolic magnetoresistive background
has been subtracted. The respective fan diagram correlates the inverse
magnetic fields at the maxima directly to different Landau levels
$n=F/B+\beta$ (Fig.~\ref{fig:WAL-SdH}d), where slope $F$ is the
oscillation frequency and offset $\beta$ is defined by the Berry
phase $\varphi_{B}=2\pi\beta$\cite{Novoselov2005-jx}. We extract
$\beta=0.52$ yielding $\varphi_{B}\approx\pi$, which is expected
for the Dirac electrons in the topological surface states. Consequently,
we can use the Onsager's relation $F=\frac{1}{2\pi}\left(\frac{\hbar c}{2\pi e}\right)\pi k_{F}^{2}=\left(\frac{\hbar c}{e}\right)n_{2D}$
to calculate the surface charge carrier concentration\cite{Ando2013-zk}
$n_{2D}=\unit[5.9\cdot10^{12}]{cm^{-2}}$, which is comparable to
the charge carrier concentration extracted from Hall measurements
(see Appendix Part~A3). Since Hall measurements probe both bulk and
surface transport, the identical charge carrier concentration from
Hall and SdH measurements at $\unit[2]{K}$ implies a vanishing bulk
background at low temperatures. The temperature and angle dependence
of the WAL and the Berry phase of the SdH oscillations are clear evidence
for a surface-dominated transport at low temperature.

\textbf{Electrical detection of SML.} BSTS with a dominant surface
transport is ideal to study the contribution to the magnetoresistance
signal due to SML. Figure~\ref{fig:SML}a shows the measurement principle
on a multi-terminal BSTS device. By applying a charge current we generate
a spin polarization ($P_{S}$) in the BSTS surface states due to SML.
This net spin polarization beneath the FM tunnel contact (Co/TiO$_{2}$)
is detected as a voltage signal depending on parallel or anti-parallel
alignment of the FM and the spin orientation. Sweeping the in-plane
magnetic field, the magnetization of the FM detector is switched with
respect to $P_{S}$ yielding a step in the voltage signal. An up-
and down-sweep of the magnetic field results in a hysteretic switching
presented in Fig.~\ref{fig:SML}c, as measured on a BSTS flake of
$\unit[70]{nm}$ thickness at $\unit[2]{K}$ by applying a DC current
of $\unit[+50]{\mu A}$.  Similarly, the spin polarization in the
BSTS surface states can be flipped by inverting the current direction
($\unit[-50]{\mu A}$ in Fig.~\ref{fig:SML}b), resulting in a reversed
spin signal (Fig.~\ref{fig:SML}d) \cite{Dankert2015-ns}.   Measuring
the full bias range, we observe a linear dependence of the spin signal
$\Delta V$ (Fig.~\ref{fig:SML}e), as expected, since the spin density
scales linearly with the current density \cite{Hong2012-zr}. However,
the strong spin-orbit coupling in the TI as well as stray fields at
the magnetic contacts can yield contributions from Rashba states,\cite{Pesin2012-uf}
spin Hall or stray Hall effects,\cite{De_Vries2015-np,Appelbaum2014-at}
mimicking a SML signal. Previously, such spurious effects have been
ruled out by several control experiments using angle,\cite{Dankert2015-ns}
gate \cite{Liu2015-tm} and carrier dependent \cite{Li2016-wd,Li2016-mn}
measurements. In particular the stray Hall effect is strongly charge
carrier type dependent, whereas the spin locking is not. Therefore,
we studied Sb$_{2}$Te$_{3}$, a known p-type TI, to confirm the sign
of the detected MR signal (see Appendix Part~A6). The observed SML
signal matches our data reported here on BSTS, as well as previous
studies on Bi$_{2}$Se$_{3}$ and BST \cite{Liu2015-tm,Dankert2015-ns,Tang2014-io,Ando2014-fh},
which confirms the signal origin stemming from the SML in the TI surface
states. Additionally, we observe an enhanced spin resistance $R_{S}=\Delta V/I$
up to $\unit[1.5]{\Omega}$ in BSTS at low temperature, which is orders
of magnitude higher than aforementioned spin Hall and stray Hall effects
\cite{Liu2015-tm}. Furthermore, such a high $R_{S}$ is also at least
one order of magnitude higher than previously reported SML results
on metallically doped TIs \cite{Li2014-jw,Dankert2015-ns}, comparable
to reports on BSTS\cite{Ando2014-fh}, and can be attributed to the
reduced number of bulk carriers at low temperature.

\begin{figure}
\begin{centering}
\includegraphics[scale=1.2]{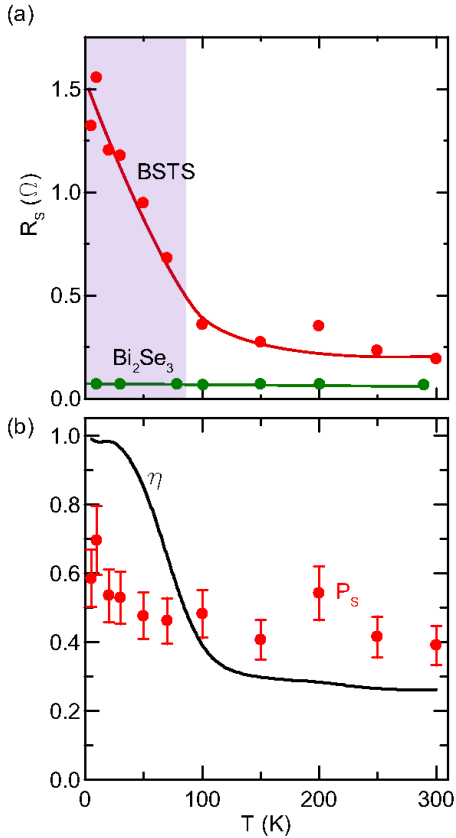}
\par\end{centering}
\caption{\textbf{Temperature dependence of SML signal. }(a)~Temperature dependence
of the spin signal amplitude $R_{S}=\Delta V/I$ for BSTS compared
with Bi$_{2}$Se$_{3}$ \cite{Dankert2015-ns} and correlated with
the channel's metal-semiconductor transition (coloured background:
SML, white background: SBC). The solid lines are guides to the eye.
(b)~Temperature dependence of the surface contribution factor $\eta=G^{S}/\left(G^{S}+G^{B}\right)$
and (c)~ the surface spin polarization $P_{S}$.\label{fig:Tdep}}
\end{figure}

In order to understand the origin of the enhancement of the signal
at low temperature, we carried out a detailed temperature dependence
measurement of the magnetoresistance. We observed a clear MR switching
up to room temperature, which decreases drastically when warming up
from $\unit[2-100]{K}$ and levels off in the SBC regime $\unit[100-300]{K}$
(Fig.~\ref{fig:Tdep}a and Appendix Fig.~A5). Previous reports on
potentiometric measurements on BSTS show a similar temperature dependence
\cite{Ando2014-fh,Ren2011-gh}, however, in a low temperature range.
In contrast, on highly doped TIs, different studies\cite{Dankert2015-ns,Mellnik2014-eb,Liu2015-tm,Jamali2015-eb,Li2014-jw,Tang2014-dq}
showed only a weak temperature dependence over large temperature ranges
(see also Appendix Fig.~A6). 

In our case, the change in the temperature dependence at around $\unit[100]{K}$
coincides with the transition from surface to SBC transport behaviour
(Fig.~\ref{fig:Device-schematic}d). Below the activation gap, the
number of bulk carriers and the inelastic scattering rate of surface
states increase exponentially with temperature \cite{Saha2014-yt,Liu2015-tm}.
This temperature dependence becomes polynomial when the temperature
exceeds the activation gap, as seen in Fig.~\ref{fig:Tdep}a \cite{Dankert2013-se,Dankert2015-ns}.
This indicates that the detected signal stems from SML in the surface
states partly suppressed by a temperature dependent surface-bulk scattering
and thermally activated bulk charge carriers \cite{Yazyev2010-kl,Hong2012-zr,De_Vries2015-np}.

Next, we fit the temperature-dependent resistance shown in Fig.~\ref{fig:Device-schematic}d
to a two-channel transport model. In this model, the thermally activated
bulk conductance is given by $G^{B}\left(T\right)=t\left(\rho_{B}\text{exp}\left(\Delta/kT\right)\right)^{-1}$,
where $k$ is the Boltzmann constant, $\rho_{B}$ is the high-temperature
bulk resistivity and $\Delta$ is the activation gap between the Fermi
level and the closest bulk states. Similarly, the surface state conductance
reads
\begin{equation}
G^{S}\left(T\right)=\begin{cases}
\left(R^{*}+BT^{4}+CT^{5}\right)^{-1} & T<T_{\text{BG}}\\
\left(R^{*}+AT\right)^{-1} & T>T_{\text{BG}}
\end{cases}.\label{eq:G-S}
\end{equation}
The surface resistance $R^{*}=R_{0}+D\text{exp}\left(\Delta/kT\right)$,
with the low temperature residual resistance $R_{0}$ and an exponential
increase with temperature of the scattering from surface to bulk states
($D$), is dominated by non-linear phonon contribution parameters
($B$ and $C$) at low temperatures (below the Bloch-Grüneisen temperature
$T_{\text{BG}}=2\hbar v_{s}k_{\text{F}}/k_{\text{B}}\approx\unit[10]{K}$)
\cite{Giraud2012-qt} and the linear electron-phonon scattering parameter
$A$ for $T>T_{\text{BG}}$. From the fitting, one may extract the
fraction of the current that flows on the surface states, $\eta=G^{S}/\left(G^{S}+G^{B}\right)$.
This fraction decreases rapidly from 100\% at low temperature to about
30\% at $\unit[100]{K}$, beyond which it remains almost constant
up to $\unit[300]{K}$ (Fig.~\ref{fig:Tdep}b). In contrast to previous
transport models\cite{Xu2014-hp}, this behaviour matches our experimentally
observed resistance change and theoretical expectations of a thermally
activated dominant bulk conduction in thin-film TIs well. 

Finally, we can estimate the spin resistance $R_{s}$, due to SML
induced by a current $I_{S}=\eta I$ yielding the surface spin polarization
$P_{S}$, which can be expressed as 
\begin{equation}
R_{s}=\frac{\Delta V}{I}=\eta R_{BT}P_{S}P_{FM},\label{eq:Rs}
\end{equation}
 where $P_{FM}$ is the polarization of the FM detector and $R_{BT}$
is the ballistic resistance\cite{Hong2012-zr}. The former has been
found in previous studies to have an upper limit of $P_{FM}=\unit[20]{\%}$
for our Co/TiO$_{2}$ contacts\cite{Kamalakar2015-ou,Dankert2015-ns}.
The ballistic conductance $\frac{1}{R_{BT}}$ equals the quantum
of conductance $\frac{q^{2}}{h}$ multiplied by the number of propagating
modes $\frac{k_{\text{F}}W}{\pi}$, where $W$ is the width of the
conductance channel. The Fermi wave number $k_{\text{F}}$ can be
derived from the 2D charge carrier concentration $n_{2D}$ as $k_{\text{F}}=\sqrt{4\pi n_{2D}}$.\cite{Dankert2015-ns}
Using Eq.~(\ref{eq:Rs}) and the surface charge carrier concentrations
measured in SdH and Hall measurements, we calculate a current-induced
surface spin polarization $P_{S}\approx0.4-0.6$, which is almost
constant within the error margins over the full temperature range
(Fig.~\ref{fig:Tdep}b) and twice as high as previously reported
\cite{Hsieh2009-ee,Li2014-jw,Yazyev2010-kl,Dankert2015-ns}. The deviation
from the optimal $P_{S}=1$ originates from spin-orbit entanglement
in the topological surface states \cite{Yazyev2010-kl}, as well as
the possible occurrence of non-topological surface states at the Fermi
level \cite{Bahramy2012-bf,Bianchi2010-jf}.  Optimising the crystal
growth and doping could yield a lower bulk conduction and hence a
high surface spin signal under ambient conditions, which would present
a serious competition for ferromagnetic contacts in spintronic applications.

In summary, we have presented the electrical detection of spin polarized
surface currents due to SML on BSTS by FM tunnel contacts over a broad
temperature range up to $\unit[300]{K}$. High quality semiconducting
BSTS crystals and FM tunnel contacts allowed the observation of a
spin signal of up to $\unit[1.5]{\Omega}$, which is at least one
order of magnitude higher than previously reported results on metallically
doped TIs in the same temperature range \cite{Dankert2015-ns,Li2016-wd}.
The large magnitude of the signal and its sign, combined with quantum
transport measurements prove clearly the SML originating from surface
states at low temperature.  Furthermore, we observe a strong temperature
dependence of the SML signal up to $\unit[100]{K}$, thereafter remaining
constant up to room temperature. A two-channel transport model considering
thermally activated bulk carriers and surface-to-bulk scattering confirms
an almost temperature independent surface spin polarisation $P_{S}$.
This elucidates the influence of the bulk conduction and scattering
mechanisms that suppress the detected SML signal, whilst still observable
up to room temperature\cite{Dankert2015-ns}. 

\section*{}

$\vphantom{1cm}$

\noindent \textbf{Acknowledgement$\quad$}The authors acknowledge
financial supports from the European Union\textquoteright s Horizon
2020 research and innovation programme under grant agreement No 696656
(Graphene Flagship), EU FlagEra project (from Swedish Research council
VR No. 2015-06813), Swedish Research Council VR project grants (No.
2016-03658), Graphene center and the AoA Nano program at Chalmers
University of Technology. The authors also acknowledge the support
of colleagues at the Quantum Device Physics Laboratory and Nanofabrication
Laboratory at Chalmers University of Technology.

\pagebreak{}

\setcounter{figure}{0}
\setcounter{section}{0}
\makeatletter  
\renewcommand{\thesection}{A\@arabic\c@section}
\renewcommand{\thefigure}{A\@arabic\c@figure}
\renewcommand{\thetable}{A\@arabic\c@table}    
\onecolumngrid 

\part*{Appendix}

\section{Fabrication and characterization}

\begin{figure}[H]
\begin{centering}
\includegraphics[scale=1.2]{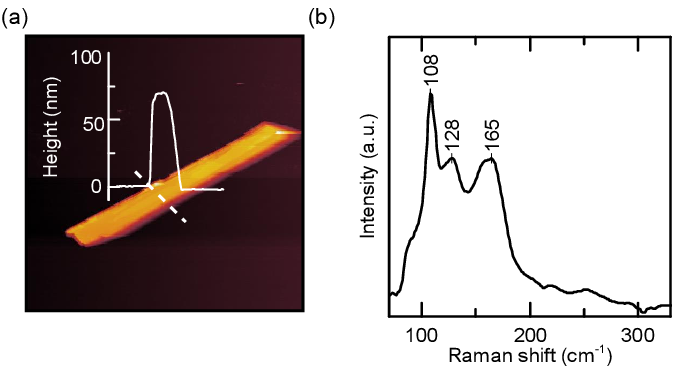} 
\par\end{centering}
\caption{\textbf{\small{}Characterization of BSTS.}{\small{} }(a)~Atomic force
micrograph of a $\unit[70]{nm}$ thick BSTS flake. Inset: Height profile
along the dashed line. (e)~Raman spectrum of a $\unit[70]{nm}$ BSTS
flake.}

\label{AFM-Raman} 
\end{figure}

\textbf{ }The Bi$_{1.5}$Sb$_{0.5}$Te$_{1.7}$Se$_{1.3}$ flakes
were exfoliated from a bulk crystal, using the conventional Scotch
tape cleavage technique, onto a clean SiO$_{2}$ ($\unit[285]{nm}$)/highly
doped n-type Si substrate. The crystal was obtained from Miracrys,
grown from a melt using a high vertical Bridgeman method \cite{Dankert2015-ns}.
The flakes were identified using optical microscopy and the flake
thickness, uniformity, and material quality was analyzed with atomic
force microscopy and Raman spectroscopy, respectively (Fig.~\ref{AFM-Raman}a
and \ref{AFM-Raman}b). This revealed homogeneous $\unit[70]{nm}$
thick flakes with widths of about $\unit[1]{\mu m}$ and a characteristic
Raman spectrum for BSTS \cite{Tu2014-qt}. Electrodes were patterned
by electron beam lithography. The contact deposition was performed
in an ultra-high vacuum electron beam evaporator after an \emph{in
situ} surface cleaning using low power argon ion plasma for $\unit[10]{s}$.
Electrodes with widths $\unit[0.3-1]{\mu m}$ and channel length of
$\unit[0.2-1]{\mu m}$ are used. As contact material, we used TiO$_{2}$/Co
for the detection of spin-momentum locking. \foreignlanguage{english}{The
$\unit[\approx1.5]{nm}$ TiO$_{2}$ tunnel barrier was deposited by
electron beam evaporation and \emph{in situ} oxidation using a pure
oxygen atmosphere. } \textbf{ }The BSTS devices were measured with
a Quantum Design Physical Property Measurement System~(PPMS) with
resistivity option using direct current (DC).

\section{Characterization of ferromagnetic tunnel junctions}

\begin{figure}[H]
\begin{centering}
\includegraphics[scale=1.2]{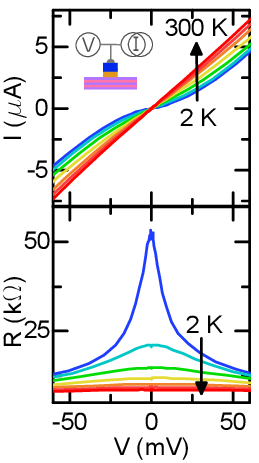} 
\par\end{centering}
\caption{\textbf{\small{}Characterization of ferromagnetic tunnel junctions.}{\small{}
}Temperature dependence of the contact current-voltage characteristic
(top panel) and resistance curves (bottom panel). The nonlinearity
indicates a good tunnel barrier with a weak temperature dependence.}

\label{TunnelCont-1} 
\end{figure}

The tunnelling properties of the ferromagnetic~(FM) tunnel contacts
are characterized in a three-terminal measurement configuration (Fig.~\ref{TunnelCont-1}).
The current-voltage characteristics exhibit a non-linear behaviour
typical for tunnelling transport. Furthermore, the temperature dependence
shows an increase of the resistance by only a factor of 5 with decrease
in temperature indicating a good quality TiO$_{2}$ tunnel barrier
on the BSTS flakes \cite{Jonsson-Akerman2000-ib}.

\section{Temperature dependence of carrier mobility and concentration}

\begin{figure}[H]
\begin{centering}
\includegraphics[scale=1.2]{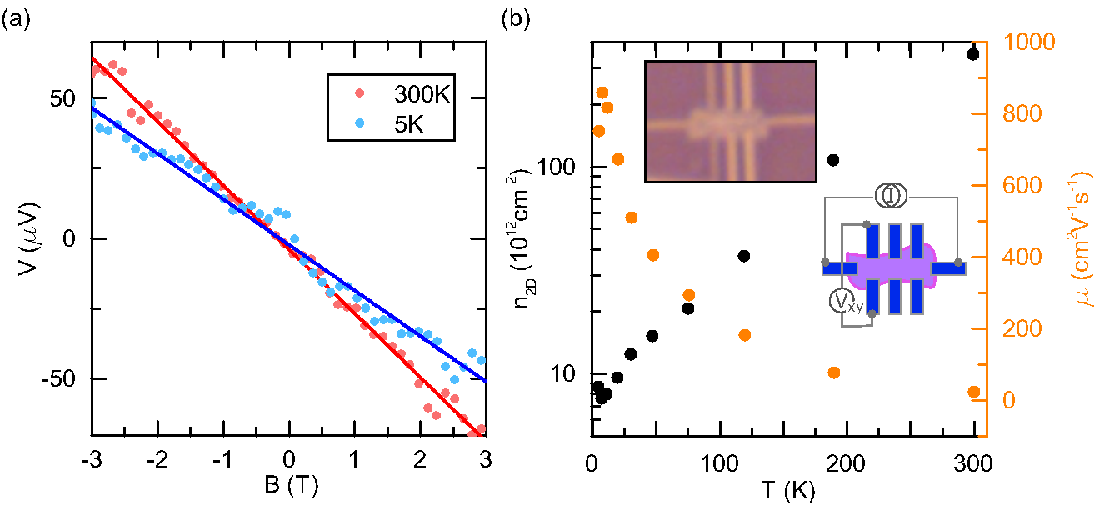} 
\par\end{centering}
\caption{\textbf{\small{}Hall measurement in BSTS.}{\small{} }(a)~The Hall
voltage measured in BSTS using a bias current of $\unit[10]{\mu A}$
at room temperature and $\unit[5]{K}$. (b)~Temperature dependence
of the Hall charge carrier concentration $n$ and mobility $\mu$.
Insets: Optical microscope image of Hall device and measurement schematic.}

\label{Hall-1} 
\end{figure}

Hall measurements were performed on a Bi$_{1.5}$Sb$_{0.5}$Te$_{1.7}$Se$_{1.3}$
channel between $\unit[2]{K}$ and $\unit[300]{K}$(Fig.~\ref{Hall-1}).
By applying a perpendicular magnetic field the Hall voltage across
the channel width is detected, allowing us to extract the 2D charge
carrier concentration $n_{2D}$ and mobility $\mu$. Fig.~\ref{Hall-1}
shows a low $n_{2D}$ at $\unit[2]{K}$, comparable to the surface
charge carrier concentration extracted from Shubnikov-de Haas measurements
(see main manuscript), and increasing with temperature, due to thermally
excited carriers from the bulk. The mobility at room temperature is
about $\unit[25]{cm^{2}\left(Vs\right)^{-1}}$, and increases to $\unit[825]{cm^{2}\left(Vs\right)^{-1}}$
at $\unit[2]{K}$. This behaviour matches well with an intrinsically
low doped semiconductor freeze-out in the bulk at low temperatures
with a remaining charge carrier concentration stemming from the surface
states at $\unit[2]{K}$.

\section{Weak anti-localization}

\begin{figure}[H]
\begin{centering}
\includegraphics[scale=1.2]{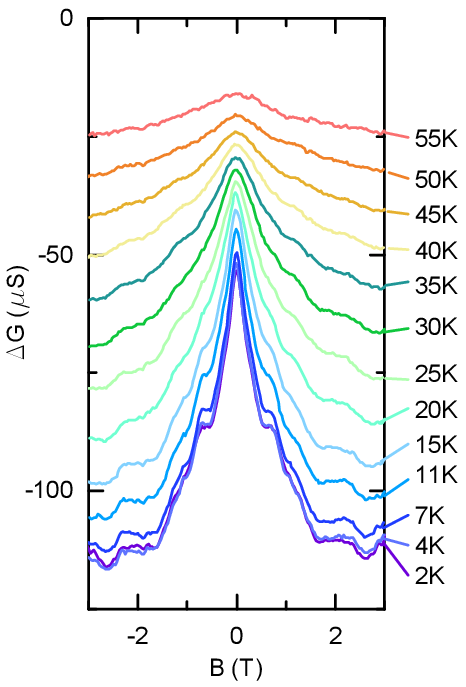} 
\par\end{centering}
\caption{\textbf{Weak anti-localization (WAL). }Magnetoconductance measurements
of BSTS with applied perpendicular magnetic field showing WAL up to
$\unit[55]{K}$ measured at $\unit[I=5]{\mu A}$. A background without
WAL signal measured at $\unit[70]{K}$ has been subtracted.}

\label{WAL-Tdep} 
\end{figure}

The quantum transport properties of BSTS were measured by sweeping
an out-of-plane magnetic field while measuring the lateral channel
resistance. We observe a correction of the magnetoresistance as expected
for strong SO coupled materials resulting in weak anti-localization
up to $\unit[55]{K}$ (Fig.~\ref{WAL-Tdep}). 

\section{Temperature dependence of SML signals}

\begin{figure}[H]
\begin{centering}
\includegraphics[scale=1.2]{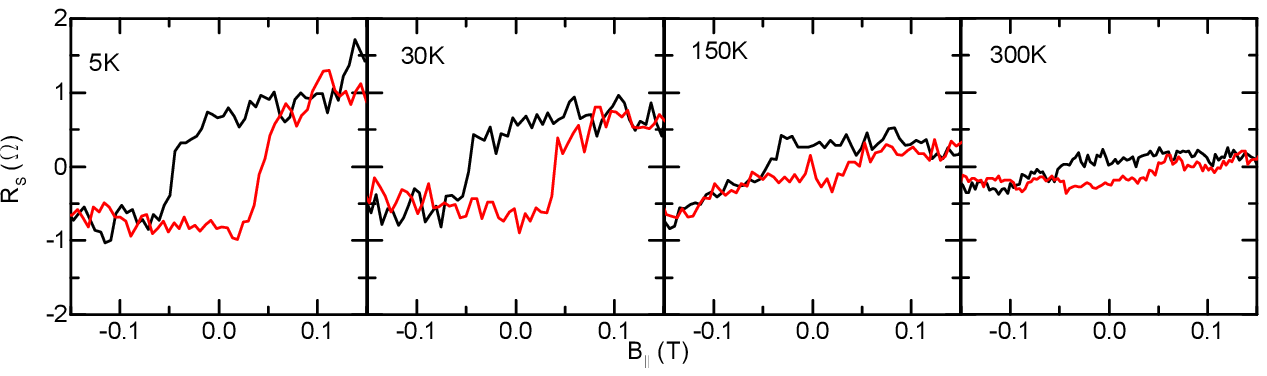} 
\par\end{centering}
\caption{\textbf{Temperature dependence of SML signal.} Spin signal measured
at different temperatures at $\unit[20]{\mu A}$ for BSTS. }

\label{SML-Tdep} 
\end{figure}

\section{Comparison of SML signals}

\begin{figure}[H]
\begin{centering}
\includegraphics[scale=1.4]{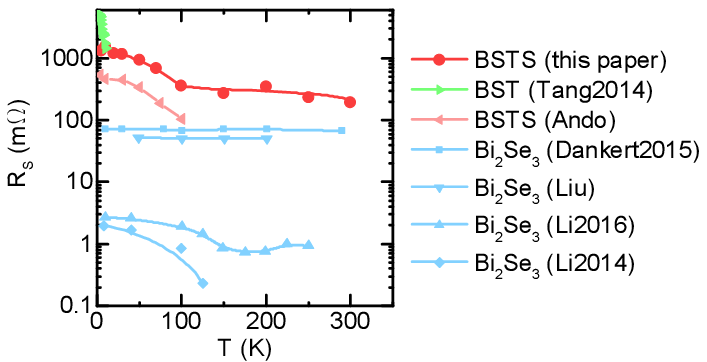} 
\par\end{centering}
\caption{\textbf{Comparison of various datasets of SML measurements from literature.}
The temperature dependences of the spin resistance observed in several
TI compounds published since the first report of electrical SML detection
by Li2014\cite{Li2014-jw}, various reports on Bi$_{2}$Se$_{3}$
(Liu\cite{Liu2015-tm}, Dankert2015\cite{Dankert2015-ns}, Li2016\cite{Li2016-wd}),
in (Bi$_{0.53}$Sb$_{0.47}$)$_{2}$Te$_{3}$(Tang2014\cite{Tang2014-io})
and Bi$_{1.5}$Sb$_{0.5}$Te$_{1.7}$Se$_{1.3}$(this paper and Ando\cite{Ando2014-fh}).}

\label{SMLcomp-1} 
\end{figure}

The amplitude and the temperature dependence of the SML signal in
our BSTS devices are compared in Fig.~\ref{SMLcomp-1} with literature
values since the first report of electrical SML detection by Li2014\cite{Li2014-jw},
various studies on Bi$_{2}$Se$_{3}$ (Liu\cite{Liu2015-tm}, Dankert2015\cite{Dankert2015-ns},
Li2016\cite{Li2016-wd}), in (Bi$_{0.53}$Sb$_{0.47}$)$_{2}$Te$_{3}$(Tang2014\cite{Tang2014-io})
and Bi$_{1.5}$Sb$_{0.5}$Te$_{1.7}$Se$_{1.3}$(this paper and Ando\cite{Ando2014-fh}).
The SML signal voltage is compared as spin resistance. Our SML signal
amplitude is at least one order of magnitude higher than reported
in metallically doped Bi$_{2}$Se$_{3}$. The temperature dependence
of our BSTS sample spans the full temperature range of $\unit[2-300]{K}$.
Only two other studies on metallic Bi$_{2}$Se$_{3}$ cover a similar
range\cite{Dankert2015-ns,Li2016-wd}. Below $\unit[100]{K}$, the
strong signal increase in our measurement matches well the results
on semiconducting BSTS\cite{Ando2014-fh} and BST\cite{Tang2014-io}.
All studies on metallic Bi$_{2}$Se$_{3}$ show a weak temperature
dependence, whereas the variation presented in studies with low detected
spin potential splitting\cite{Li2014-jw,Li2016-wd} may be due to
other influences, and the temperature dependence is also negligible
compared to BSTS.

\section{Spin-momentum locking in p-type TI Sb$_{\ensuremath{2}}$Te$_{\ensuremath{3}}$}

\begin{figure}[H]
\begin{centering}
\includegraphics[scale=1.2]{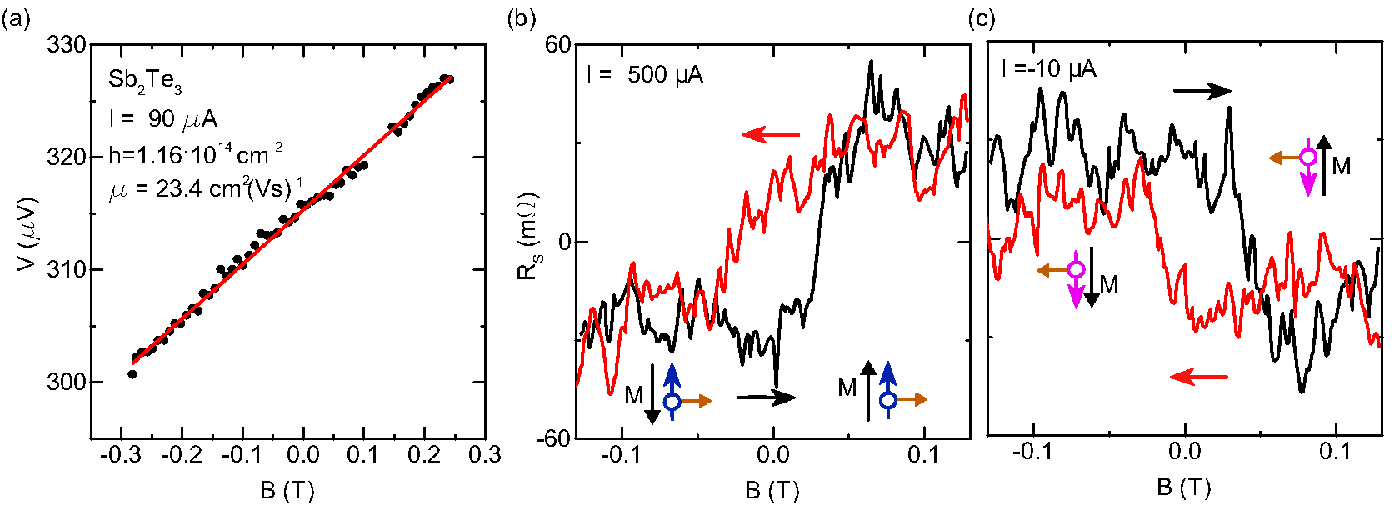} 
\par\end{centering}
\caption{\textbf{Spin-momentum locking in p-type Sb$_{\ensuremath{2}}$Te$_{\ensuremath{3}}$.
}(a)~ Hall measurement showing p-type doping. SML measured at opposite
bias polarities of (b)~$\unit[500]{\mu A}$ and (c)~$\unit[-10]{\mu A}$.}

\label{SbTe} 
\end{figure}

The stray fields at the magnetic contacts can yield stray Hall effects,
which can appear similar to the SML signal. However, such spurious
signals are strongly charge carrier dependent and change sign when
switching from electron to  hole transport. In contrast, the spin
polarisation of the SML does not depend on the charge carriers, since
the locking as well as the momentum are opposite for holes and electrons.
We studied Sb$_{2}$Te$_{3}$, a known p-type TI, to verify the origin
of the measured signal. Figure~\ref{SbTe}a shows the Hall measurement
in Sb$_{2}$Te$_{3}$, which exhibits a majority hole carrier concentration
of $p=\unit[1.16\cdot10^{14}]{cm^{-2}}$. The observed SML signal
for positive (Fig.~\ref{SbTe}b) and negative (Fig.~\ref{SbTe}c)
bias current shows the typical hysteretic switching and matches our
data reported here on BSTS, as well as previous studies on Bi$_{2}$Se$_{3}$
and BST \cite{Liu2015-tm,Dankert2015-ns,Tang2014-io,Ando2014-fh},
which confirms the signal origin stemming from the SML in the TI surface
states.

\end{document}